\documentclass[9pt,twocolumn,twoside]{osajnl}
\journal{optica} % Choose journal (ao, aop, josaa, josab, ol, optica, pr)

\setboolean{shortarticle}{false}
\title{Time-reversed photoacoustic guided time-reversed ultrasonically encoded optical focusing}

\author[1]{Juze Zhang}
\author[1]{Zijian Gao}
\author[1]{Xiaopeng Yu}
\author[1]{Peng Ge}
\author[1]{Feng Gao}
\author[1,*]{Fei Gao}
\affil[1]{ Hybrid Imaging System Laboratory, Shanghai Engineering Research Center of Intelligent Vision and Imaging, School of Information Science and Technology, ShanghaiTech University, Shanghai 201210, China}

\affil[*]{Corresponding author: gaofei@shanghaitech.edu.cn}

\begin{abstract}
	Deep-tissue optical imaging is a longstanding challenge limited by scattering. Both optical imaging and treatment can benefit from focusing light in deep tissue beyond one transport mean free path. Wavefront shaping based on time-reversed ultrasonically encoded (TRUE) optical focusing utilizes ultrasound focus, which is much less scattered than light in biological tissues as the 'guide star'. However, the traditional TRUE is limited by the ultrasound focusing area and pressure tagging efficiency, especially in acoustically heterogeneous medium. Even the improved version of iterative TRUE comes at a large time consumption, which limits the application of TRUE. To address this problem, we proposed a method called time-reversed photoacoustic wave guided time-reversed ultrasonically encoded (TRPA-TRUE) optical focusing by integrating ultrasonic focus guided by time-reversing PA signals, and the ultrasound modulation of diffused coherent light with optical phase conjugation (OPC), achieving dynamic focusing of light into scattering medium. Simulation results show that the focusing accuracy of the proposed method has been significantly improved compared with conventional TRUE, which is more suitable for practical applications that suffers severe acoustic distortion, e.g. transcranial optical focusing. 
\end{abstract}

\setboolean{displaycopyright}{false}

\begin{document}

\maketitle

\section{Introduction}

Optical imaging in deep tissues is an open challenge in biomedical imaging research. The focus of light in deep tissues can not only enhance the output signal intensity of optical imaging systems (such as fluorescence imaging), but also improve imaging sensitivity and depth. On the other hand, it can also be directly applied to optogenetics, photodynamic therapy, laser ablation and other fields that require light energy concentration \cite{grier2003revolution, boyden2005millisecond, yanik2004functional}.

However, one longstanding challenge in optical imaging is scattering. Strong scattering causes the wavefront to be distorted as they pass through the tissue. Meanwhile, the photons are absorbed by the tissue during propagation, which greatly limits the penetration of light into deeper biological tissues. When a beam of light enters the tissue, photons will bounce many times, and its movement will become random after transport mean free path \cite{wang2012deep}. Until recently, most optical techniques used for imaging inside tissues mainly rely on only un-scattered (i.e. ballistic) photons such as confocal microscopy, two-photon microscopy and optical coherence tomography (OCT) \cite{ntziachristos2010going}. Therefore, the ability to focus light inside the scattering medium can revolutionize the deep-tissue non-invasive fluorescence microscope, optical tweezers, optogenetics, microsurgery, and phototherapy, covering many areas of biophotonic applications\cite{liu2017focusing}.

Wavefront shaping may be the best method to address this problem by shaping the wavefront of the incident light to counteract the scattering \cite{kubby2019wavefront}. Three main classes of wavefront shaping approaches are being actively developed, including model-based wavefront shaping \cite{thendiyammal2020model}, feedback-based wavefront shaping \cite{mosk2012controlling} and optical time reversal/optical phase conjugation (OPC) \cite{yanik2004time, yaqoob2008optical}.

Model-based wavefront shaping depends on the numerical computation using a digital model of the sample. The microscopic refractive index model of the sample, which serves as the input for the calculations, is obtained from the image data itself \cite{thendiyammal2020model}. Feedback-based wavefront shaping depends on the detection of the intensity feedback \cite{vellekoop2007focusing} at a desired location to find the optimal wavefront maximizing that feedback signal, by using some advanced optimization algorithm such as genetic algorithm \cite{vellekoop2007focusing}, and even deep learning \cite{yang2020deep}. This feedback signal is often obtained by using a camera to measure the scattered light, but it is unrealistic to place a camera inside biological tissue. Lai et al \cite{lai2015photoacoustically}. proposed to use the photoacoustic signal as a feedback signal to optimize the wavefront of light, which makes the feedback-based method suitable for biological tissues. Despite this, the process of focusing consumes a lot of time to optimize, which limits its wide applications.

\begin{figure*}[htbp]
	\centering\includegraphics[width= 9cm]{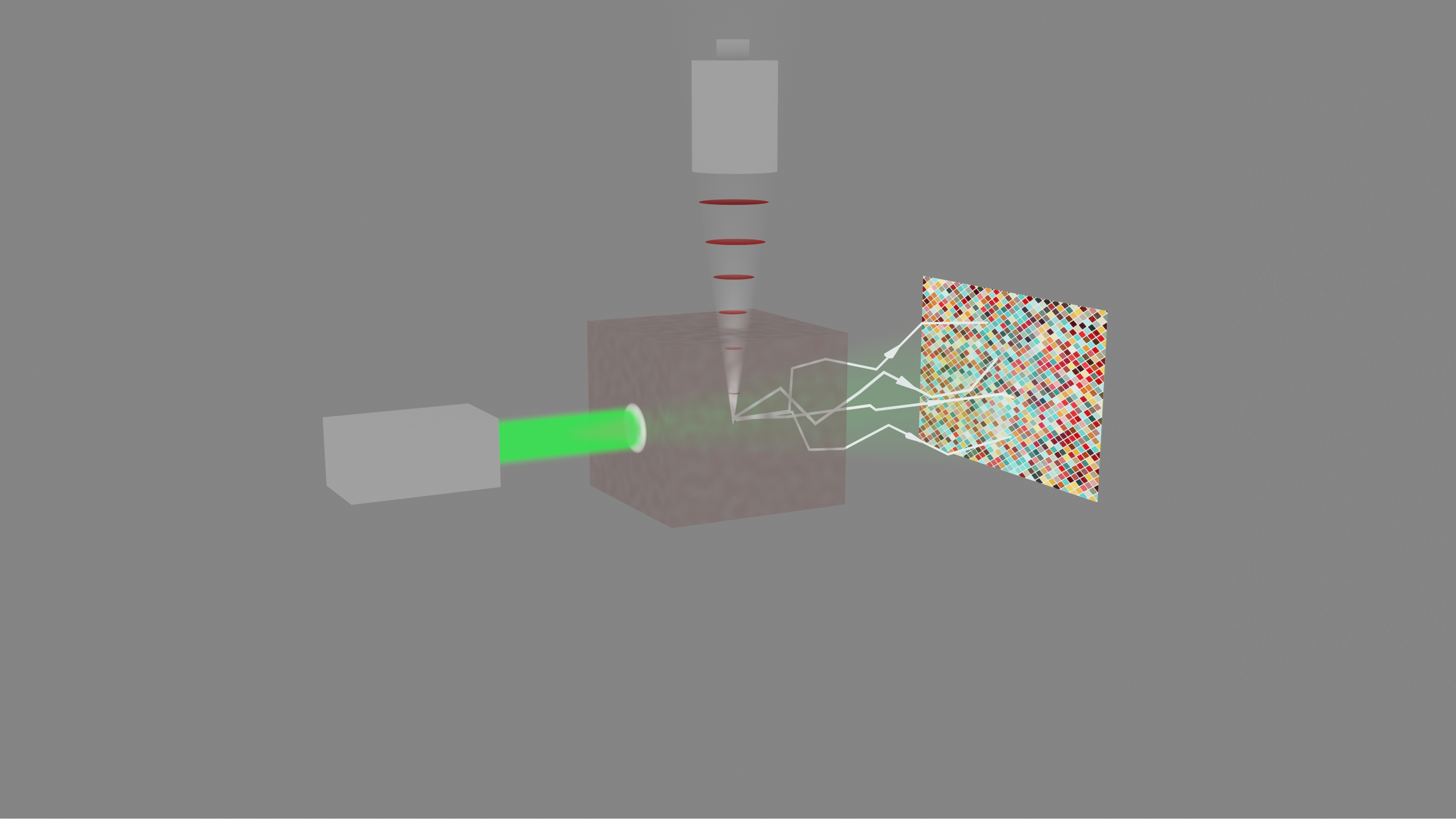}
	\centering\includegraphics[width= 9cm]{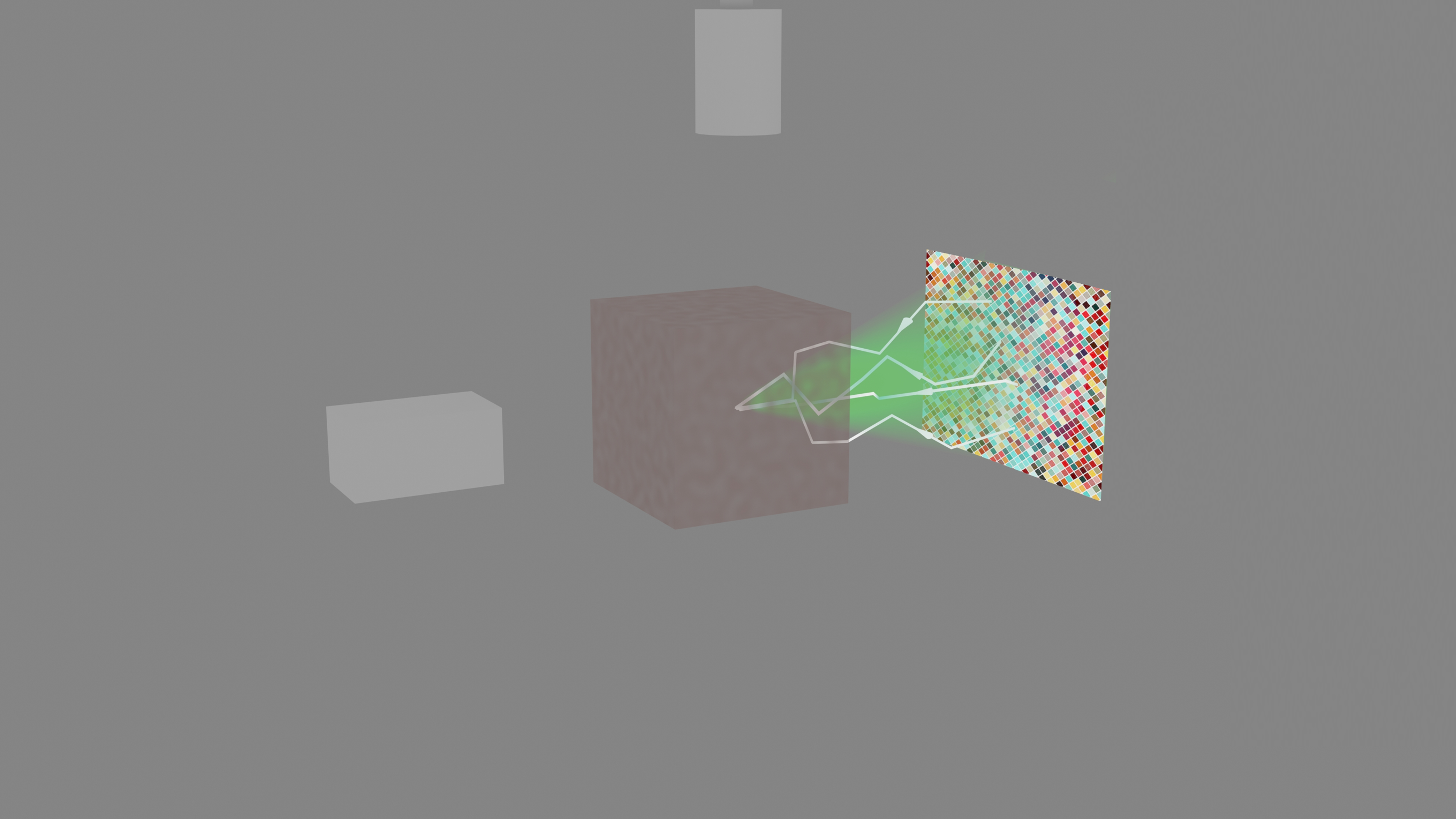}
	\caption{ Simplified schematic diagram of principle for TRUE optical focusing. (a) Initially, a wide laser beam or a narrow laser beam is incident on the scattering medium. After entering the scattering medium, billions of photons undergo some scattering events, and the path of the photons becomes random. When the scattered photon passes through the position under insonification, part of the photons are tagged by the ultrasound, resulting in a frequency shift. The white solid line represents the photon path of the tagged part, and finally reaches the SLM and then is recorded. (b) Through the holographic method, we can calculate the phase of the photon in the tagged part, take the conjugate of the phase and load it on the SLM. As a result, the photons will follow their path and return to the region under insonification, just like going back in time.
	}
	\label{TRUE}
\end{figure*}

Alternatively, in OPC, the optimal wavefront is obtained from a single measurement of the scattered photons propagating from a source located behind or inside the turbid medium. Subsequently, a focus is formed by playing back the conjugate of this field using a phase conjugate mirror. Similar to feedback-based wavefront shaping, OPC requests a 'guide star' in the scattering medium to guide the wavefront focusing \cite{horstmeyer2015guidestar, vellekoop2008demixing}. To address this, Xu et al \cite{xu2011time} proposed a scheme termed time-reversed ultrasonically encoded (TRUE) optical focusing, which uses ultrasound focus as the 'guide star'. Figure~\ref{TRUE} illustrates the process of TRUE. Unlike light, ultrasound is much less scattered in biological tissues, which can create a virtual source of light frequency shifted by the acousto-optic effect as the 'guide star'. The ultrasound emitted by a focused ultrasound transducer modulates scattered photons (ultrasonically encoded) (Fig.~\ref{TRUE}a). As a result, the encoded photons will have a frequency shift to form a 'guide star' in the ultrasound focus area. As a 'new' light source, the 'guide star' forms an interference pattern with the reference light and is recorded by the holographic recording medium. When the hologram is reproduced, the phase-conjugated object light will return to the ultrasonic modulation area through time reversal, forming an optical focus inside the scattering medium (Fig.~\ref{TRUE}b).

The resolution of the early TRUE technology is determined by the size of the ultrasound focus (25-50$\mu$m) \cite{horstmeyer2015guidestar}. However, the focusing ability of ultrasound is essential to ensure sufficient localized energy deposition and sufficient number of photons modulated in the target area (see Section 2, Methods). When the target area is located behind an acoustically heterogeneous layer (such as an irregular layer of adipose tissue, or bone), the acoustic focusing performance becomes the main bottleneck limiting its wide application. One possible solution that has been proposed to solve the TRUE resolution problem is iterative TRUE \cite{si2012breaking, suzuki2014continuous, ruan2014iterative}, that is, the feedback loop between the conjugate field and the detection field can reduce the focus size by three times validated through experiments. Nevertheless, the time consuming of iterative schemes will increase the complexity of the system.

Recently, photoacoustic (PA) generation in the target have been proposed as an alternative, e.g. Fink et al. have successfully verified the feasibility of time-reversal of PA signals \cite{bossy2006time}. Moreover, our previous works have also demonstrated the potential of ultrasonically focus guided by time-reversing the transcranial PA signals of an optically selective target in a nonselective background \cite{zhang2020photoacoustic}.

In this paper, we propose a method by hybrid integrating photoacoustic, ultrasonic time-reversal, and OPC to achieve real-time ultrafast 2D/3D optical focusing in both optically and acoustically heterogeneous medium with one snapshot.

\section{Method}

\begin{figure*}[ht!]
	\centering\includegraphics[scale = 0.35]{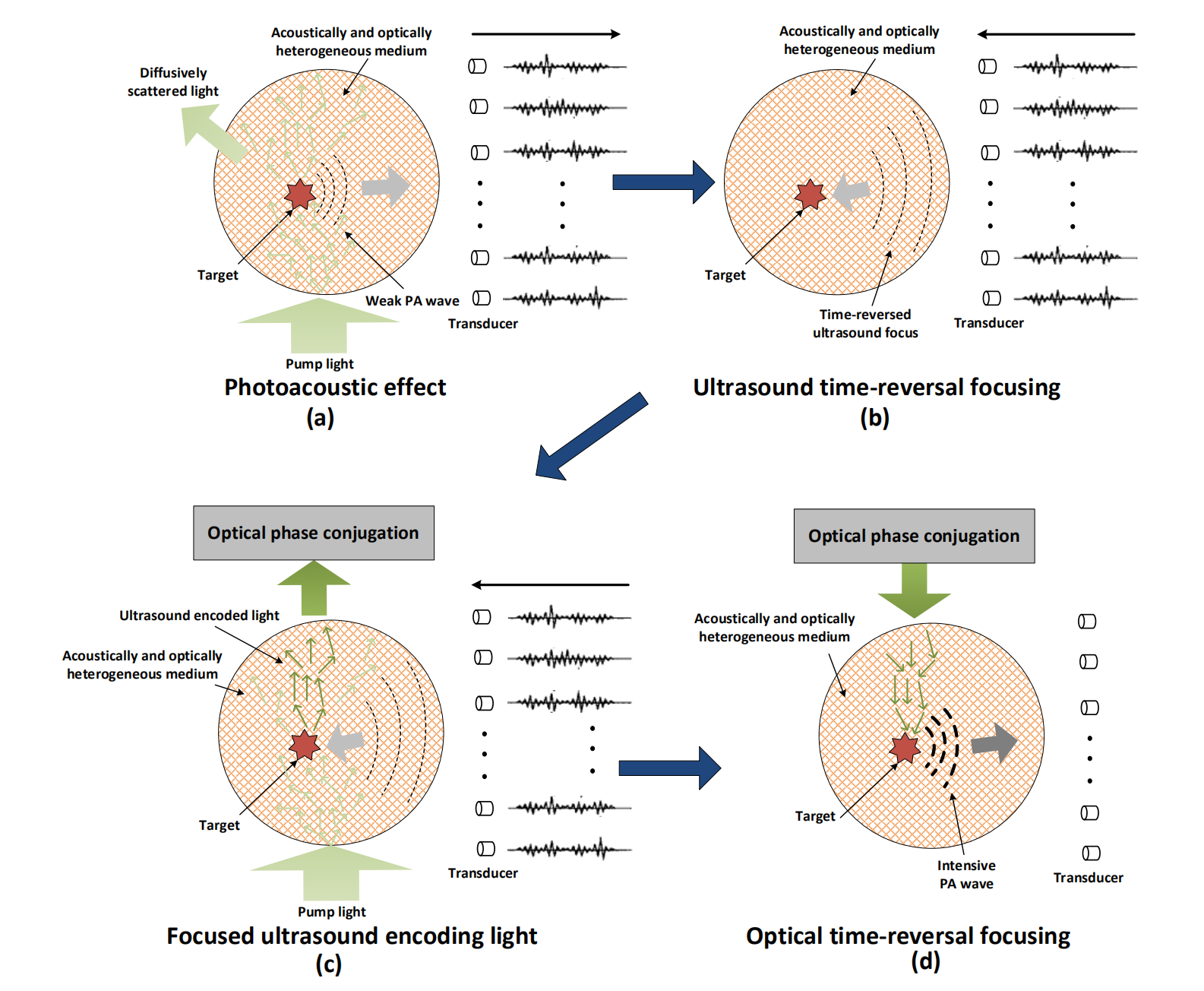}
	\caption{ Simplified schematic diagram of principle for TRPA-TRUE optical focusing. (a) First, an unfocused light from a laser source ( $\lambda$ =532 nm) is illuminated into the medium and optically absorbing regions act as acoustic sources through the photoacoustic effect, which generates photoacoustic wave and then be detected. (b) Subsequently, we perform time reversal operation on the received PA signal to achieve ultrasound focusing. (c) When the time-reversed PA signal reaches the optical absorption regions, a part of the photons will be tagged in this snapshot in the optical absorption area and then reach the OPC. (d) Finally, we perform TURE operation to complete TRPA-TURE.
	}
	\label{TRPA-TRUE}
\end{figure*}

Our proposed method, called time-reversed photoacoustic wave guide time-reversed ultrasonically encoded (TRPA-TRUE) optical focusing, integrates ultrasonic focus guided by time-reversing PA signals, and the ultrasonic modulation of diffused coherent light with OPC, achieving dynamic focusing of light into a scattering medium. Figure~\ref{TRPA-TRUE} illustrates the concept of our method. First, an unfocused light from a laser source ( $\lambda$=532 nm) is illuminated into the medium. Optically absorbing regions act as acoustic sources through the photoacoustic effect, which generates a photoacoustic wave detected by transducer array (see Fig.~\ref{TRPA-TRUE}a). Subsequently, we perform acoustic time reversal operation on the received PA signals to achieve ultrasound focusing inside the medium (Fig.~\ref{TRPA-TRUE}b). When the time-reversed PA signal reaches the optical absorption regions, part of the photons will be tagged in this snapshot in the optical absorption area and then reach the OPC (Fig.~\ref{TRPA-TRUE}c). Finally, we perform TURE operation to complete whole process of TRPA-TRUE, which achieves real-time ultrafast 2D/3D optical focusing in both optically and acoustically heterogeneous medium with one snapshot.

In the following sections, we will start with the acousto-optic effect and explain why the focus of ultrasound determines our TRUE effect, then we will explain the feasibility of time reversal of the PA signal. Finally, we will also perform a simulation study to verify our method.

\subsection{Acousto-Optic Interaction within Turbid Media}

There are currently three widely recognized modes to interpret the AO effect: The displacement of the scatterers in the turbid medium caused by the acoustic field \cite{leutz1995ultrasonic}, the disturbance of the refractive index of the medium caused by the mechanical strain due to the acoustic field \cite{wang2001mechanisms1}, and perturbation of the optical properties of the medium due to compression and rarefaction of the medium by the acoustic field \cite{mahan1998ultrasonic}. However, the last mechanism is very weak, so we can ignore it here \cite{wang2001mechanisms2}.

\subsubsection{Ultrasound-induced change of the refractive index}
Denote the acoustic pressure field at  position $\bf{r}$ and time $t$ as the following equation:
 \begin{equation}
 	P( \mathbf{r}, t ) = P_0 \sin( \omega_at - \mathbf{k_a} \cdot \mathbf{r})
 	\label{eq:refname1}
 \end{equation}
 where $P_0$ is the acoustic amplitude, $\omega_a $ is the acoustic angular frequency and  $\mathbf{k_a}$ is the acoustic wave vector. 
 
Acoustic waves propagating through the medium produce strain, which alters the local permittivity of medium. The adiabatic piezo-optic coefficient $\eta = \frac{\partial n}{\partial p}$ relates the change in the index of refraction $n$ to the
acoustic pressure $	P( \mathbf{r}, t )$. The refractive index change can be described as:

 \begin{equation}
	\Delta n = n_0( \frac{\partial n}{\partial p}) P( \mathbf{r}, t )
	\label{eq:refname2}
\end{equation}
where $n_0$ is the background refractive index without ultrasound modulation.

\subsubsection{Ultrasound-induced displacement of scatterers.}

To describe the locations of any particles, we model the ultrasound-induced displacement of particle at  position $\bf{r}$ and time $t$ by using the following equation:
 \begin{equation}
	A( \mathbf{r}, t ) = A_0 \mathbf{\hat{k}_a} \sin( \omega_at - \mathbf{k_a} \cdot \mathbf{r} + \Phi)
	\label{eq:refname3}
\end{equation}
where $ A_0$ is the displacement amplitude, $\mathbf{\hat{k}_a} $ is the unit vector of $\mathbf{k_a}$, and $\Phi$ is particular
phase offset between particle displacement and ultrasound pressure. Recently, Huang et al. have investigated the ultrasound–light
interaction in scattering media\cite{huang2020investigating}, and we can know from their theory that $A_0 = \frac{P_0 }{\omega_a \rho v_a}$, where $\rho$ is the density of the background medium and $ v_a$ is the velocity of acoustic wave. In their theory, 
phase offset $\Phi$ can be approximately replaced by $3 \pi/ 2$ between particle displacement and ultrasound pressure when when particles move.

Combining (\ref{eq:refname2}) and (\ref{eq:refname3}), we can easily get that when the Monte Carlo(MC) simulation is performed, the number of the ultrasound encoded photons shows positive correlation with acoustic intensity, whether due to ultrasound-induced change of the refractive index or ultrasound-induced displacement.

\subsubsection{Ultrasound tagging efficiency in scattering media}

 Another important factor, tagging efficiency, should be taken into account. Here we use $\eta$ to represent tagging efficiency and we give its mathematical expression as following:

\begin{equation}
	\eta = \frac{P_{\text{tag}}}{P_{\text{untag}}+P_{\text{tag}}}
	\label{eq:refname4}
\end{equation}

where $P_{\text{tag}}$ is the power of light that is tagged by ultrasound pressure and $P_{\text{untag}}+P_{\text{tag}}$ is the total power of light that pass through the ultrasound pressure field. It is worth noting that here we only consider the part of the light passing through the ultrasound pressure field, not all the photon power. 
Unfortunately, we did not derive the mathematical expression between the tagging efficiency and the ultrasound pressure field, so here we prove the relationship between them through simulation.

To calculate the tagging efficiency, we only need to consider power spectrum of one speckle exiting the skull and we represent it by electric field $E(t)$. We have  $E(t)=\sum_{i}E(\theta_i,t)$, where $\theta_i$ represents the $i$'th photon in this speckle. Because in MC simulation, each photon is  independent, we did not treat the photon as a wave phenomenon and ignore features such as phase and polarization. So we can calculate power spectrum by the following equation\cite{huang2020investigating}:

\begin{equation}
	\mathbb{E}\{ | \mathcal{F}[E(t)] |^2\}  = 	\mathbb{E}\{ \sum_{i} | \mathcal{F}[E(\theta_i,t)] |^2\}
	\label{eq:refname5}
\end{equation}

where $\mathbb{E}$ represents expectation and we can calculate power spectrum by $| \mathcal{F}[E(t)] |^2$. 
With the above definition, we can estimate the tagging efficiency from the overall average of the power spectrum of all speckles. But calculating all speckles seems to be too computationally expensive, so we can estimate the tagging efficiency by sampling a subset of photon through the scattering medium.

\begin{figure}[ht!]
	\centering\includegraphics[scale = 0.45]{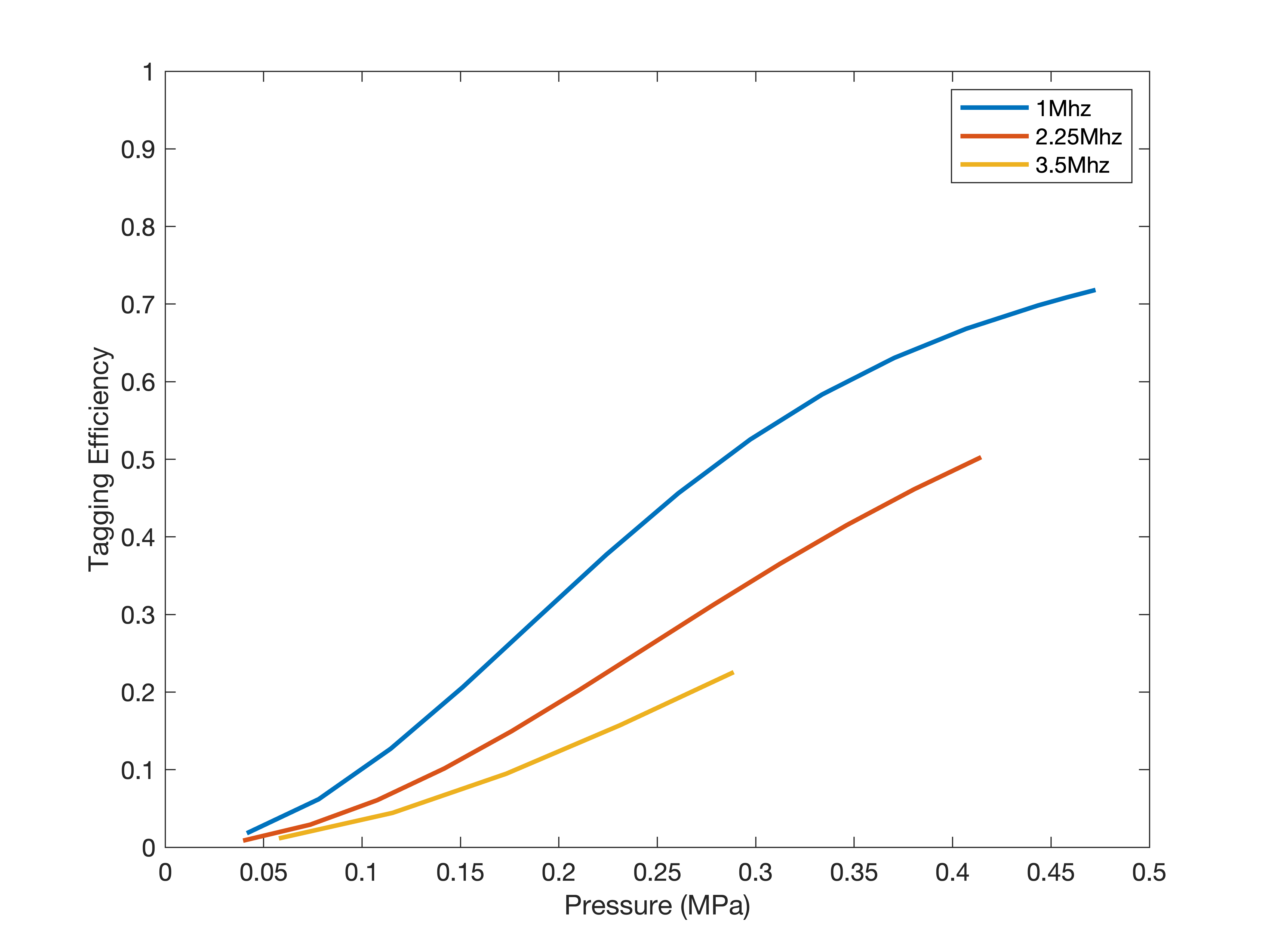}
	\caption{ Tagging efficiency with 1 MHz, 2.25MHz and 3.5MHz acoustic frequencies versus acoustic pressure.}
	\label{tagging efficiency}
\end{figure}

Thanks to Yujia Huang for opening up their ultrasound-tagging code on github and uploading their experimental data\cite{huang2020investigating}. We downloaded the code and tested it to get the relationship between the tagging efficiency and ultrasound pressure as well as the ultrasound frequency. The comparison result is shown in Fig.~\ref{tagging efficiency}.

\begin{figure*}[t!]
	\centering\includegraphics[scale = 0.18]{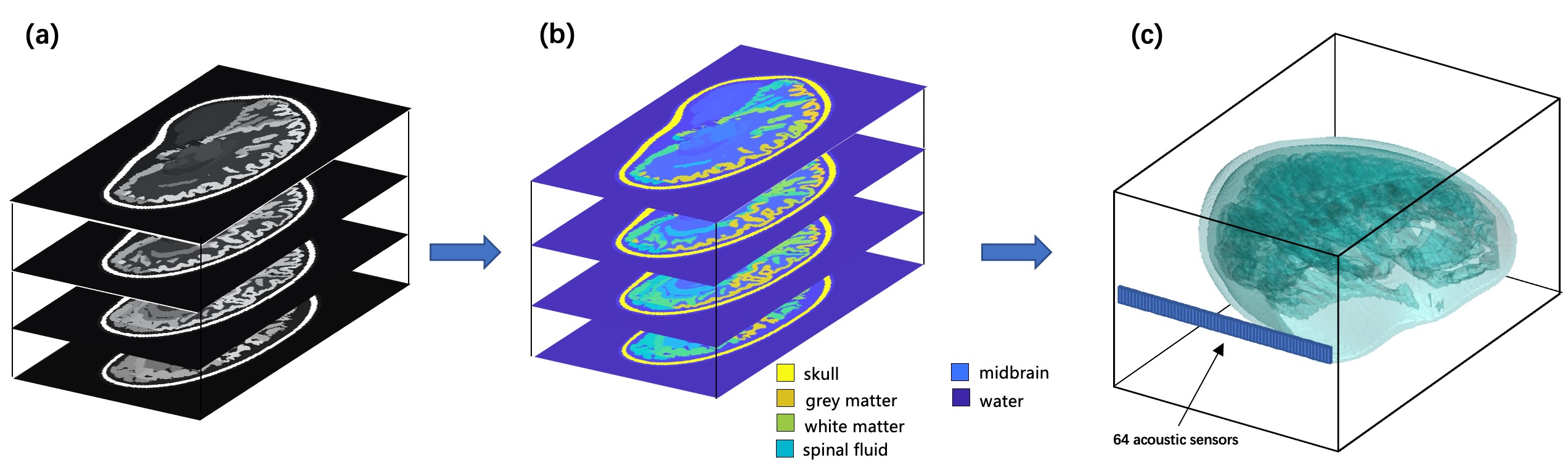}
	\caption{ The diagram of the simulation environment modeling process. (a) Get each slice from MRI data. (b) The MRI data of brain were imported into MATLAB and segmented into air, water, midbrain, white matter, grey matter, spinal fluid scalp and skull using thresholds segmentation. (c) After segmenting each slice, we stack the processed MRI images to form a 3-D model, and then load it into the k-wave simulation package.}
	\label{simulation}
\end{figure*}

Figure~\ref{tagging efficiency} shows that the higher the ultrasound pressure, the higher the ultrasound tagging efficiency. This is also consistent with our prediction that the ultrasound focus determines the TRUE focusing ability. And at the same time, we can also get that as the frequency increases, the tagging efficiency is decreasing. According to what we have learned, in order to obtain a smaller ultrasound focus area, many research groups use a higher ultrasound frequency (e.g. 50 MHz), but this will result in a decrease in the tagging efficiency of ultrasound, which increases the difficulty of detection. In addition, even if we ignore the negative effects of the reduction of ultrasound tagging, the use of higher ultrasound frequencies greatly reduces penetration depth, because high-frequency acoustic signals tend to be more easily attenuated by tissues. To sum up the above points, in order to achieve better optical focusing through the TRUE system, we first need to improve the efficiency of the ultrasound tagging ‘guide star’, and our method can greatly improve this performance.

\subsection{Time invariance of PA signals}
Optically absorbing regions act as acoustic sources through the photoacoustic effect described by the following equation\cite{tam1986applications}

\begin{equation}
\Delta p - \frac{1}{c^2}\frac{\partial^2 p}{\partial t^2} = - \frac{\beta}{c_p}\frac{\partial H}{\partial t}
\label{eq:refname6}
\end{equation}

where $p$ is the acoustic pressure, $c$ is the speed of sound, $\beta$ is the isobaric thermal expansion coefficient, $c_p$ is the specific heat, and $H$ is the volumetric density of optical energy per unit time. If $p(r,t)$ is a solution of the equation, then $p(r,-t)$ is also a solution. In other words, the equation is invariant under time reversal\cite{funke2009photoacoustic}. Thus, diverging waves emanating from a source can be reversed in time and back-propagated in space to focus at the source.

\section{Simulation of TRPA-TRUE on a Brain model}

Our method was verified by performing simulations of TRPA-TRUE within Turbid Media using a brain model (Fig.~\ref{simulation}a), which is downloaded from the Scalable Brain Atlas \cite{bakker2015scalable}. The acoustic simulations are implemented in k-space using the k-Wave toolbox of MATLAB (R2019a, Mathworks Inc., Natick, MA, USA) \cite{treeby2010k}, which provides raw PA data, and simulates the process of time-reversed sound wave propagation. In the optical simulation part, we use the Monte Carlo (MC) photon propagation model \cite{wang1995mcml}, which has been widely used in acousto-optic imaging (AOI) simulation \cite{huang2020investigating, powell2012highly, leung2010fast}, because it can directly simulate the interaction of acousto-optic in the medium. The simulation medium proprieties can be set using masked operations assigning the values from the brain model. The properties of the different intracranial propagation medium is shown in Table \ref{tab:model}.

As shown in Fig.~\ref{simulation}, the MRI data of brain were imported into MATLAB and segmented into air, water, midbrain, white matter, grey matter, spinal fluid scalp and skull using thresholds segmentation. After segmenting each slice, we stack the processed CT images to form a 3-D model, and then load it into the k-wave simulation package. Figure~\ref{simulation} illustrates the steps involved in the segmentation and acoustic brain simulation configuration process. The simulation setup in Fig.~\ref{simulation}(c) shows that 64 transducers were placed around the center of $xz$ plane to acquire the PA signals from the sample. The distance from the center to the acoustic sensors is 40 mm. The sensor is set as an ideal ultrasound detector with infinite bandwidth. The speed of sound outside the skull in the simulation is set as 1500 m/s. These sensors recorded the PA pressure of whole medium with 150 MHz sampling frequency.

\begin{table}[htbp]
	\centering
	\caption{\bf Speed, Density and Absorption of Distribution on Model According to Img-pixel-values}
	
	\setlength{\tabcolsep}{1mm}{
	\begin{tabular}{ccccc}
		\hline
		Tissue & Speed & Density & Absorption& Img-pixel\\
		  &$[m/ s]$ & ($Kg/ m^3$) & $[dB/(MHz \cdot cm)] $&-values\\
		\hline
		air                     &  343      & 1.2      & 0.0004& [0]* \\
		water               &1475     & 1000   & 0.05& [0]*\\
		midbrain         & 1546.3 &1000   & 0.6 & [21-39],[51-78]\\
		white matter & 1552.5 &1050   & 0.6 & [40-50]\\
		grey matter   & 1500     & 1100 & 0.6 & [81-220]\\
		spinal fluid    & 1475     & 1000 & 0.05 & [1-9]\\
		scalp               & 1540     & 1000 & 0.1 & [10-20]\\
		skull                & 3476     & 1979 & 2.7 & [221-255]\\
		\hline
	\end{tabular}}
	\label{tab:model}
\end{table}

\section{Results}

\subsection{TRPA Guided Focus within Distorted Medium}

\begin{figure}[ht!]
	\centering\includegraphics[scale = 0.23]{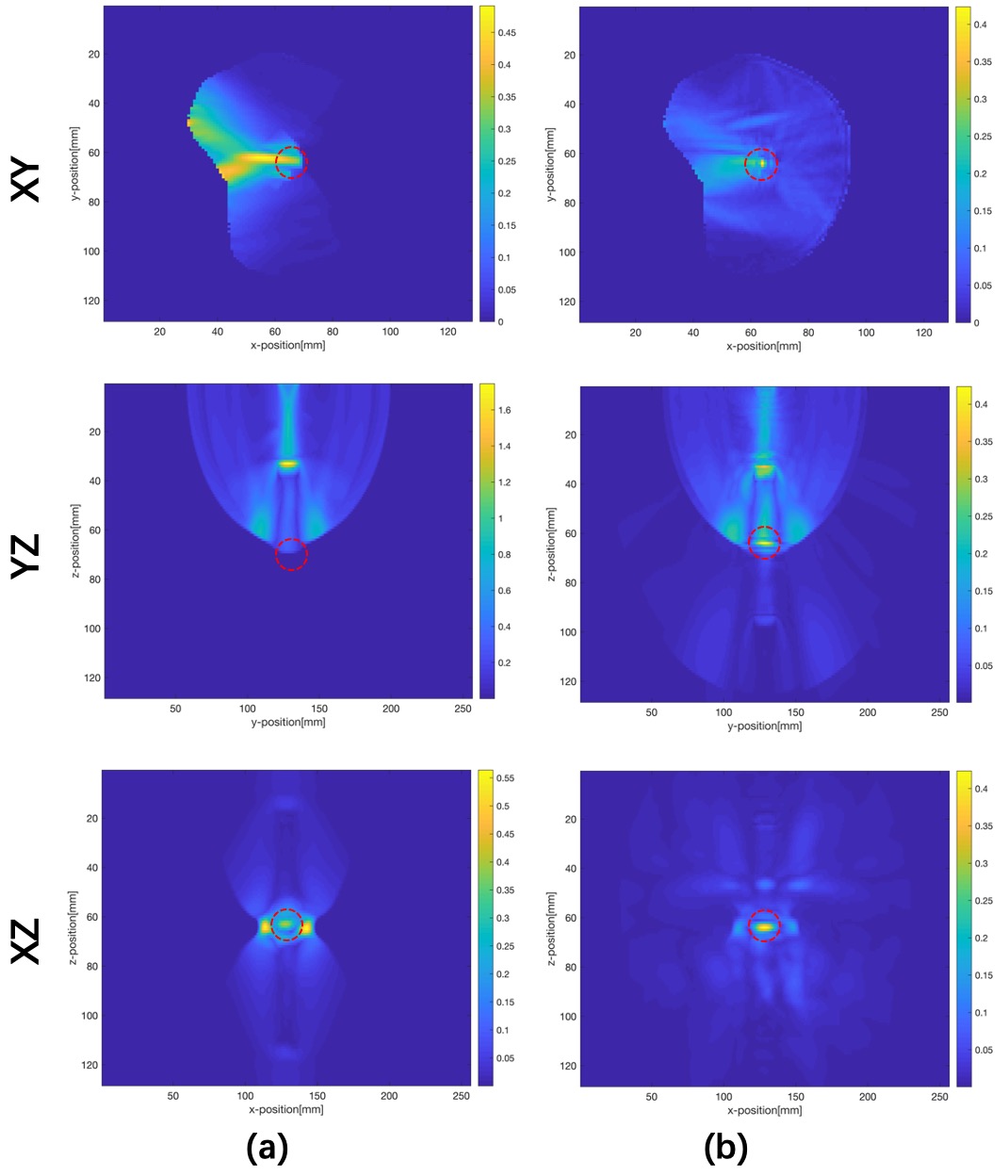}
	\caption{Simulated pressure gain distribution in the cross-section plane of xy, xz and yz. The black contours represent the shape of the skull. (a) ultrasound direct focus. (b) TRPA guided focus.}
		\label{us-result}
\end{figure}

Since the speed of light is much faster than the speed of sound, ultrasound can be regarded as static when the light passes through it, . so So here we use the maximum signal pressure recorded at each point of the ROI to evaluate the acoustic focusing ability. The PA signal was generated by the k-Wave simulation in MATLAB, followed by transcranial transmission, and received by 64 acoustic sensors. We obtained the focus of the sound waves at the initial position with and without TR. An absorber with size of 2 mm and absorption coefficient of 5 $cm^{-1}$ is placed in the center of the ROI to generate a PA signal. Experimental results are shown in Fig.~\ref{us-result}.  Our simulation environment is carried out in a 3D configuration. In order to get the acoustic focus in the space more intuitively, we intercepted three different planes of $xy$, $xz$ and $xz$ for comparison. Using the focused transducer, it resulted in an ultrasound focus at the intended location shown in Fig.~\ref{us-result}(a). We found that ultrasound is focused at the intended location, but there were many “hot spots” outside the discs with a similar or even higher pressure gain than the value inside focus. Local “hotspots” at the edge of the disc that extended out into the surrounding tissue may lead to generate unwilling ‘guide star’ that affects the accuracy of the tagging. In comparison, using TRPA guided focus (Fig.~\ref{us-result}(b)) resulted in an ultrasound focusing at the intended location with much less “hotspots” on the disc outside.

\begin{figure*}[ht!]
	\centering\includegraphics[scale = 0.235]{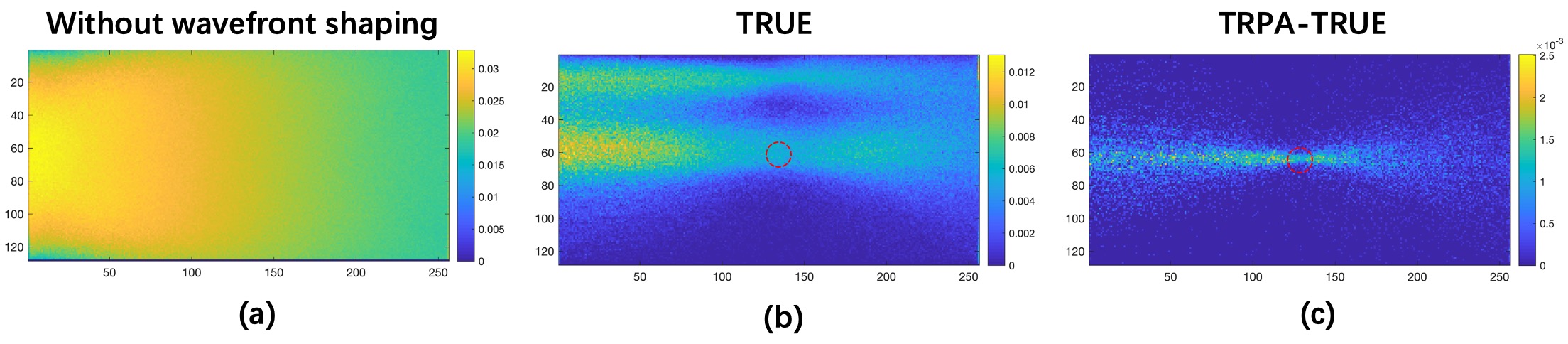}
	\caption{Simulated result of mean intensity fluence images computed along the z-axis. (a) Without wavefront shaping. (b) TRUE. (c)TRAP-TRUE. }
	\label{photo-result}
\end{figure*}

\subsection{TRPA-TRUE Optical Focus within Distorted Medium}

Figure~\ref{photo-result}(a) shows the mean intensity projection of the 3D stack acquired without wavefront shaping. We intercepted three cross-sections along y-axis to more intuitively see the focus of light in space. It can be clearly seen from Fig.~\ref{photo-result}(a) that the intensity of the flux decreases rapidly as the distance from the scattering layer increases.

Figure~\ref{photo-result}(b) shows the mean intensity projection of the fluence image after applying the wavefront shaping obtained from traditional TRUE method. Here we can clearly see that TRUE greatly improves the light focusing effect compared to that without wavefront shaping. However, around the middle of the z-axis (i.e. position of ultrasonic 'guide star'), the focus of the light seems a bit blurred with expanded focal size and one extra unwanted focal spot on top of it. More severely, the highest intensity is around the surface of the tissue, rather than at the focal spot It can also be seen that due to the low signal-to-noise ratio, the intensity does not follow a monotonic change at these depths.

Figure~\ref{photo-result}(c) shows the mean intensity projection of the fluence image after applying wavefront shaping obtained from TRPA-TRUE method. We use the time-reversed PA signal as an excitation of ultrasound modulation to tag the propagating photons. In contrast to the traditional TRUE method, the fluence heat map shows a much smaller focus area with higher intensity projection at the target position.

\section{Discussions And Conclusion}

In this work, we demonstrated TRPA-TRUE by integrating photoacoustic, ultrasonic time-reversal, and OPC, for the first time,  achieving ultrafast 2D/3D optical focusing in both optically and acoustically heterogeneous medium with one snapshot. Using the MC simulation and k-Wave toolbox, we demonstrated the ability to enhance focus intensity with higher SNR through numerical brain model, which has not been previously reported for any TRUE system. Although time-reversed photoacoustic signals has shown the feasibility to provide ultrasound focus for acousto-optic interaction in this work, for in vivo human brain, the presence of the skull makes it challenging to generate intracranial photoacoustic signals due to the strong light scattering and attenuation of the skull, which is an open challenge to be solved.

\section{Funding}
This research was funded by Natural Science Foundation
of Shanghai (18ZR1425000), and National Natural Science
Foundation of China (61805139).

\section{Acknowledgments}

The MRI data is downloaded from Neuromorphometrics, Inc\cite{bakker2015scalable} and the ultrasound pressure data is downloaded on github from Yujia Huang. Thanks for their
support.

% Bibliography
\bibliography{ref}

% Full bibliography added automatically for Optics Letters submissions; the following line will simply be ignored if submitting to other journals.
% Note that this extra page will not count against page length
\bibliographyfullrefs{ref}

%Manual citation list
%\begin{thebibliography}{1}
%\bibitem{Zhang:14}
%Y.~Zhang, S.~Qiao, L.~Sun, Q.~W. Shi, W.~Huang, %L.~Li, and Z.~Yang,
 % \enquote{Photoinduced active terahertz metamaterials with nanostructured
  %vanadium dioxide film deposited by sol-gel method,} Opt. Express \textbf{22},
  %11070--11078 (2014).
%\end{thebibliography}

% Please include bios and photos of all authors for aop articles
\ifthenelse{\equal{\journalref}{aop}}{%
\section*{Author Biographies}
\begingroup
\setlength\intextsep{0pt}
\begin{minipage}[t][6.3cm][t]{1.0\textwidth} % Adjust height [6.3cm] as required for separation of bio photos.
  \begin{wrapfigure}{L}{0.25\textwidth}
    \includegraphics[width=0.25\textwidth]{john_smith.eps}
  \end{wrapfigure}
  \noindent
  {\bfseries John Smith} received his BSc (Mathematics) in 2000 from The University of Maryland. His research interests include lasers and optics.
\end{minipage}
\begin{minipage}{1.0\textwidth}
  \begin{wrapfigure}{L}{0.25\textwidth}
    \includegraphics[width=0.25\textwidth]{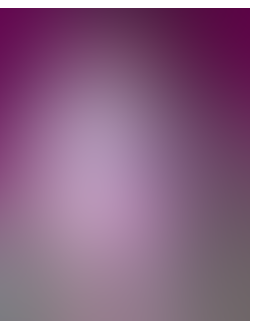}
  \end{wrapfigure}
  \noindent
  {\bfseries Alice Smith} also received her BSc (Mathematics) in 2000 from The University of Maryland. Her research interests also include lasers and optics.
\end{minipage}
\endgroup
}{}

\end{document}